\documentstyle[12pt,epsf,epsfig]{article}
\begin{document}
\begin{titlepage}
\pagestyle{empty}
\baselineskip=21pt
\rightline{hep-ph/0209005}
\vskip 0.25in
\begin{center}
{\large{\bf 
Lepton-Flavor Violating Decay of Tau Lepton \\
in the Supersymmetric Seesaw Model
}}
\end{center}
\begin{center}
\vskip 0.25in
{\bf Junji Hisano}
\vskip 0.15in
{\it
{ICRR, University of Tokyo, Kashiwa 277-8582, Japan }
}
\vskip 0.25in
{\bf Abstract}
 
Now the bounds on the lepton-flavor violating (LFV) decay modes of tau
lepton are being improved by the Belle experiment in the KEK B
factory. In this paper the LFV decay of tau lepton is discussed in the
supersymmetric seesaw model, using the current neutrino oscillation
data.
\end{center}
\baselineskip=18pt \noindent

\vfill
\leftline{September 2002}
\end{titlepage}
\baselineskip=18pt


Lepton-flavor Violation (LFV) in the neutrino sector is observed by
the oscillation experiments. The $\nu_\mu$-$\nu_\tau$ mixing was first
discovered in the atmospheric neutrino \cite{skatm}, and is being
confirmed furthermore by the K2K experiments \cite{k2k}. Also, the
neutrino oscillation in the solar neutrino is confirmed by the SNO
experiment \cite{sno}, and it is considered to comes from
$\nu_e$-$\nu_\mu$ mixing.  The SNO and the superKamiokande experiments
pin down the solar neutrino problem to the LMA solution almost.  This
will be confirmed soon by the Kamland experiments. Now, the neutrino
oscillation experiments use human-made neutrinos, and we are going to
next stage for LFV in the neutrino sector. Now we have a big question,
that is, {\it how large LFV in the charge lepton sector is}.  The
neutrino oscillation comes from the small neutrino masses, and we need
to know the origin of the neutrino mass in order to answer the
question.

The seesaw model \cite{seesaw}, in which the right-handed neutrinos
are introduced, predicts the finite, but tiny neutrino masses in a
simple way. In this model, the Yukawa coupling and Majorana mass terms
are lepton-flavor violating. In addition to the small neutrino mass,
this model is well-motivated from other theoretical points of
view. First, in the SO(10) GUT, the unification of matter is realized
in an economical and elegant way. Quarks and leptons in one generation
are embedded in a {\bf 16} dimensional multiplet. Here, existence of the
right-handed neutrinos is required. Second is the baryon asymmetry in
Universe.  Since the sphareron effect washed out the $B+L$ number in
the early universe, the $B-L$ generation is required. The Majorana
masses of the the right-handed neutrino violate the lepton number, and
thus, the leptogenesis \cite{lepto} induced by the right-handed
neutrinos is a good candidate of the baryogenesis.

It is considered that the seesaw model should be supersymmetrized
since this model has a big hierarchy between the weak scale and the
right-handed neutrino mass scale. The supersymmetry has a big role for
LFV in the charged lepton sector. The radiative correction by the
Yukawa coupling of the right-handed neutrino may generate the sizable
LFV slepton masses \cite{bm}, and the predicted $Br(\mu\rightarrow
e \gamma)$ or $Br(\tau\rightarrow \mu \gamma)$ may be in the reach
of the future experiments \cite{lfv} \cite{ehrs}. 

In this paper we concentrate on the lepton-flavor violating tau decay
in the supersymmetric seesaw model. Now the bounds on the
lepton-flavor violating tau decay modes are being improved by the Belle
experiment in the KEK B factory. The new  bounds by the Belle
experiment are following;
\begin{eqnarray}
Br(\tau\rightarrow \mu \gamma) &<& 5\times 10^{-7},
\\
Br(\tau\rightarrow e^-e^+e^-) &<& 7.8\times 10^{-7},
\\
Br(\tau\rightarrow \mu^-\mu^+\mu^-) &<& 8.7\times 10^{-7},
\\
Br(\tau\rightarrow e^-\mu^+\mu^-) &<& 7.7\times 10^{-7},
\\
Br(\tau\rightarrow \mu^-e^+e^-) &<& 3.4\times 10^{-7},
\\
Br(\tau\rightarrow e^-K^0) &<& 5.8\times 10^{-7},
\\
Br(\tau\rightarrow \mu^-K^0) &<& 5.3\times 10^{-7},
\end{eqnarray}
for the processes with $\Delta L_\tau=1$ and $\Delta L_{\mu/e}=1$
\cite{taumug,taulfv}. The Belle experiment is expected to  
obtain about 300$fb^{-1}$ data two year after. If they can keep the 
present situation, the bound for $Br(\tau\rightarrow
\mu \gamma)$ will be improved to be $6\times 10^{-8}$. Also, if 
the super B factory starts, it can reach to $2\times 10^{-8}$. Thus,
now it is important to discuss impact of the LFV decay of tau lepton 
on the supersymmetric seesaw model. 

First, we review the supersymmetric seesaw model briefly. The superpotential 
of the lepton sector is given as 
\begin{eqnarray}
W = N^{c}_i (Y_\nu)_{ij} L_j H_2
  + E^{c}_i (Y_e)_{ij}  L_j H_1 
  + \frac{1}{2}{N^c}_i (M_N)_{ij} N^c_j ,
\end{eqnarray}
where the indexes $i,j$ run over three generations and ${(M_N)}_{ij}$
is the heavy singlet-neutrino mass matrix. While the light-neutrino mass matrix
is given as 
\begin{eqnarray}
({\cal M_\nu})_{ij} &=&
\sum_k \frac{(Y_\nu)_{ki}(Y_\nu)_{kj}}
            {{M}_{N_k}} \langle H_2 \rangle^2
\label{lightMnu}
\end{eqnarray}
from this superpotential, the radiative LFV slepton masses induced by
the Yukawa coupling of the right-handed neutrino are approximately
proportional to an Hermitian matrix $H$,\footnote{
Here, we do not include the other LFV sources, such as in SUSY GUT
\cite{sglfv}.  However, typical SUSY GUTs predict smaller
$Br(\tau\rightarrow \mu \gamma)$ or $Br(\tau\rightarrow e \gamma)$ ,
which is suppressed by the KM mixing. Thus, our result for the LFV tau decay is
expected not to be changed even if effect of the SUSY GUT is included.}
\begin{eqnarray}
H_{ij}
&=&
\sum_k 
{(Y_\nu^\dagger)_{ki}}
{(Y_\nu)_{kj}}
\log\frac{M_G}{{M}_{N_k}} , 
\end{eqnarray}
where $M_G$ is the GUT scale.

Now we can parameterize the seesaw model by $({\cal M_\nu})$ and $H$.
The seesaw model has 18 physical degrees of freedom in addition to 3
charged lepton masses, those are 6 eigenvalues, 6 mixing angles, 4
mixing CP phases, and 2 Majorana CP phases for the Yukawa coupling and the
Majorana mass matrices. $H$ has 3 eigenvalues, 3 mixing angles, and 3
phases, while the light-neutrino mass matrix has 3 mass eigenvalues, 3
mixing angles, 1 mixing CP phase, and 2 Majorana CP phases. Thus, the
degrees of freedom in the seesaw model match with those in $({\cal
M_\nu})$ and $H$ \cite{di}. Unfortunately, the parameter space in
$({\cal M_\nu})$ and $H$ is wider than that in the seesaw model,
however, the parametrization works well to realize the seesaw model
from the low-energy data including the neutrino-oscillation data
\cite{ehrs}. As will be shown explicitly later, the fact we can take
$({\cal M_\nu})$ and $H$ almost independently means that the charged
LFV search, such as $\mu\rightarrow e \gamma$ or $\tau \rightarrow
\mu\gamma$, will give information independent of the neutrino
oscillation for the seesaw model.

Now we show how large the branching ratios of the LFV tau decay modes  can
be, using our parameterization. Here, we adopt two ansatz for $H$.
First is
\begin{eqnarray}
H_1=\left(\begin{array}{ccc}
a & 0 & 0 \\
0 & b  & d  \\
0 &  d^\dagger & c
\end{array} \right) .
\label{H1}
\end{eqnarray}
where $a,b,c$ are real and positive, and $d$ is a complex
number. Here, $H_{12}$ and $H_{12}H_{32}$ are zero. If either of them is of the
order of 1, $Br(\mu\rightarrow e \gamma)$ is much larger than the
experimental bound. This ansatz may predict large $Br(\tau\rightarrow
\mu\gamma)$ if $H_{23}$ is large.  Second is 
\begin{eqnarray}
H_2=\left(\begin{array}{ccc} a & 0 & d \\ 0 & b & 0 \\ d^\dagger & 0 &
c
\end{array} \right) \, .
\label{H2}
\end{eqnarray}
This may predict large $Br(\tau\rightarrow e \gamma)$ when $H_{13}$ is
large. From a viewpoint of model-building, $H_1$ may be more favored
than $H_2$. If the right-handed neutrino mass matrix ${M}$ is
proportional to {\bf 1}, $H_{23}$ in $H$ may be comparable to $H_{33}$
and $H_{22}$ since $\tan^2\theta_{23}\sim1$ in the MNS matrix.  Also,
we may need some conspiracy between the Yukawa coupling and the
Majorana masses of the right-handed neutrino in order to realize the
large mixing angles of the atmospheric and solar neutrino oscillation
in $H_2$. However, both possibilities are not rejected
phenomenologically at present, and thus, we explore the both cases.

In order to reconstruct the seesaw model, the light-neutrino mass
matrix also has to be fixed. We use $\Delta m^2_{32}=3\times 10^{-3}$
eV$^2,$ $\Delta m^2_{21}=4.5\times 10^{-5}$ eV$^2,$
$\tan^2\theta_{23}=1$ and $\tan^2\theta_{12}=0.4$ corresponding to the
LMA solution. Since the bound on the angle $\theta_{13}$ is quite
stringent, our results depend very weakly on its actual value. We fix
$\sin\theta_{13}=0.1$ and the CP phases in the MNS matrix
$\delta=\pi/2$. We study both the normal and the inverse hierarchy for
the light-neutrino mass spectrum, since the neutrino oscillations do not
discriminate between these two cases. As input parameters, we then
have the lightest effective neutrino mass $m_1$ (or $m_3$ for
inversely ordered neutrinos), which we generate in the range
$(10^{-4}-0.3)$ eV, the two low scale Majorana phases $\phi_{1,2}$ and
the matrix $H$, which we generate randomly.

In Fig.~1 we present Br($\tau\rightarrow \mu \gamma$) for the ansatz
$H_1$, assuming the normal or the inverted hierarchy for the
light-neutrino mass spectrum. The horizontal axis is the lighter stau
mass $m_{\tilde{\tau}_1}$, and the other supersymmetry-breaking
parameters are determined by choosing the SU(2) gaugino mass to be
200~GeV, $A_0=0$, $\mu>0$, and $\tan\beta=10$. We see from these
figures that behaviors of the branching ratios are similar for the
normal and inverted hierarchies for the light-neutrino mass spectrum.
In our parametrization the branching ratio is determined mainly by $H$
and the sparticle mass spectrum. In this figure the maximum value of
the prediction is fixed by the perturbation condition, and it is
larger than the current experimental bound. We sample the parameters
in $H_1$ randomly in the range $10^{-2}<a,b,c,|d|<10$, with
distributions that are flat on a logarithmic scale, and require the
Yukawa coupling-squared to be smaller than $4 \pi$, so that $Y_\nu$
remains perturbative up to $M_{G}$.

In Fig.~2, we present Br($\tau\rightarrow e \gamma$) for the ansatz
$H_2$, assuming the normal or the inverted hierarchy for the
light-neutrino mass spectrum. The input parameters are the same in
Fig.~1.  The significant difference between the normal and inverted
hierarchies for the light-neutrino mass spectrum does not exit, and
the maximum values of the prediction is determined by the perturbation
condition, again.  This means that both searches for $\tau\rightarrow
\mu \gamma$ and $\tau\rightarrow e \gamma$ are important and we have
possibility to observe either of them in future experiments.

Fig.~1 and Fig.~2 show that the prediction of the LFV tau decay modes may be
almost independent of the light-neutrino mass matrix. The search for
the LFV tau decay is one of important pieces to reconstruct the
seesaw model.

\begin{figure}[htbp]
\centerline{\epsfxsize = 0.5\textwidth \epsffile{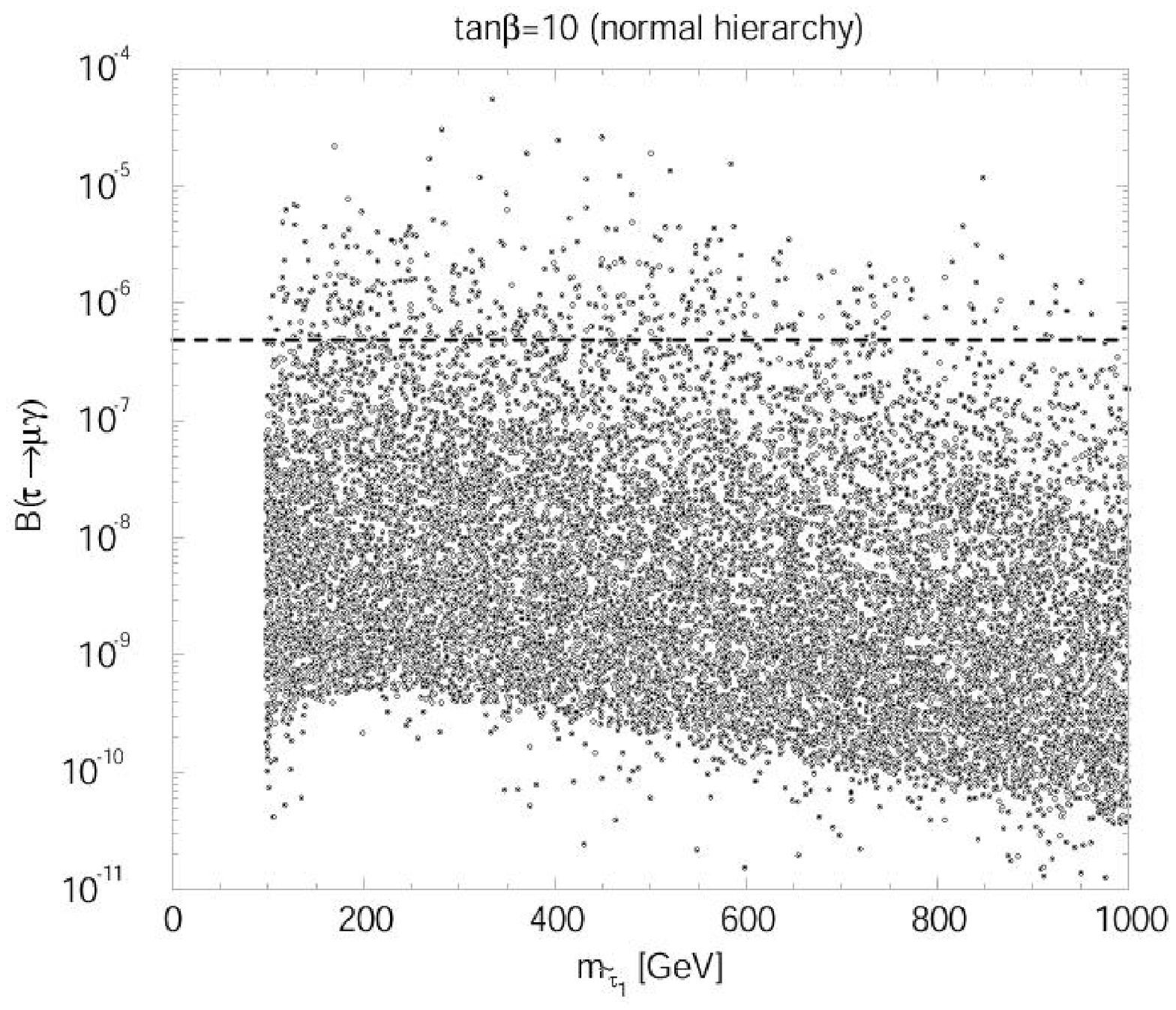} 
\hfill \epsfxsize = 0.5\textwidth \epsffile{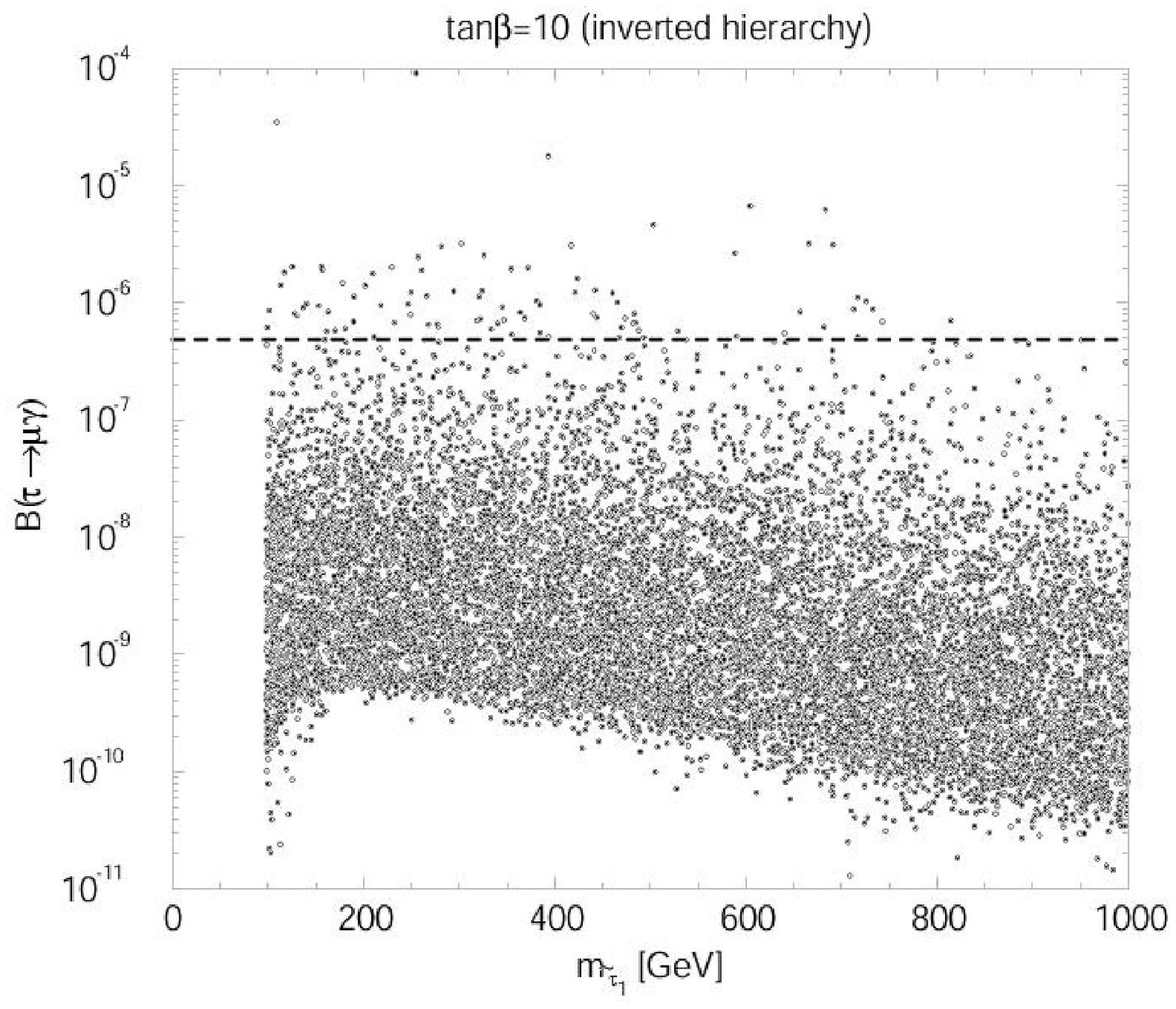} }
\caption{\it 
Scatter plot of Br$(\tau\to \mu \gamma)$ against the lighter stau mass for 
the
ansatz $H_1$. We take the SU(2) gaugino mass to be 200~GeV, $A_0=0$,
$\mu>0$, and $\tan\beta=10$. We assume both the normal
and inverted hierarchies for the light-neutrino mass spectrum.
\vspace*{0.5cm}}
\end{figure}

\begin{figure}[htbp]
\centerline{\epsfxsize = 0.5\textwidth \epsffile{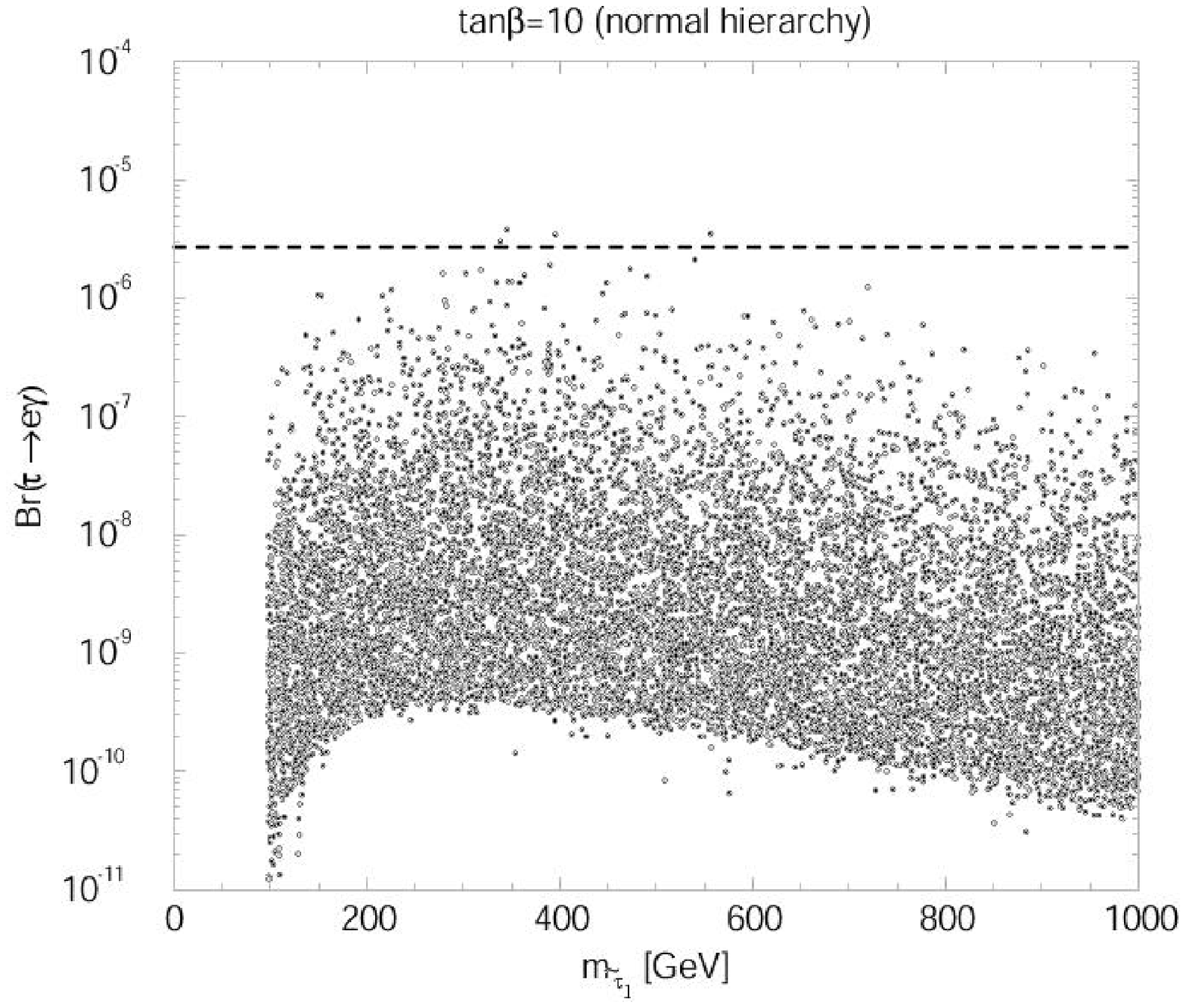} 
\hfill \epsfxsize = 0.5\textwidth \epsffile{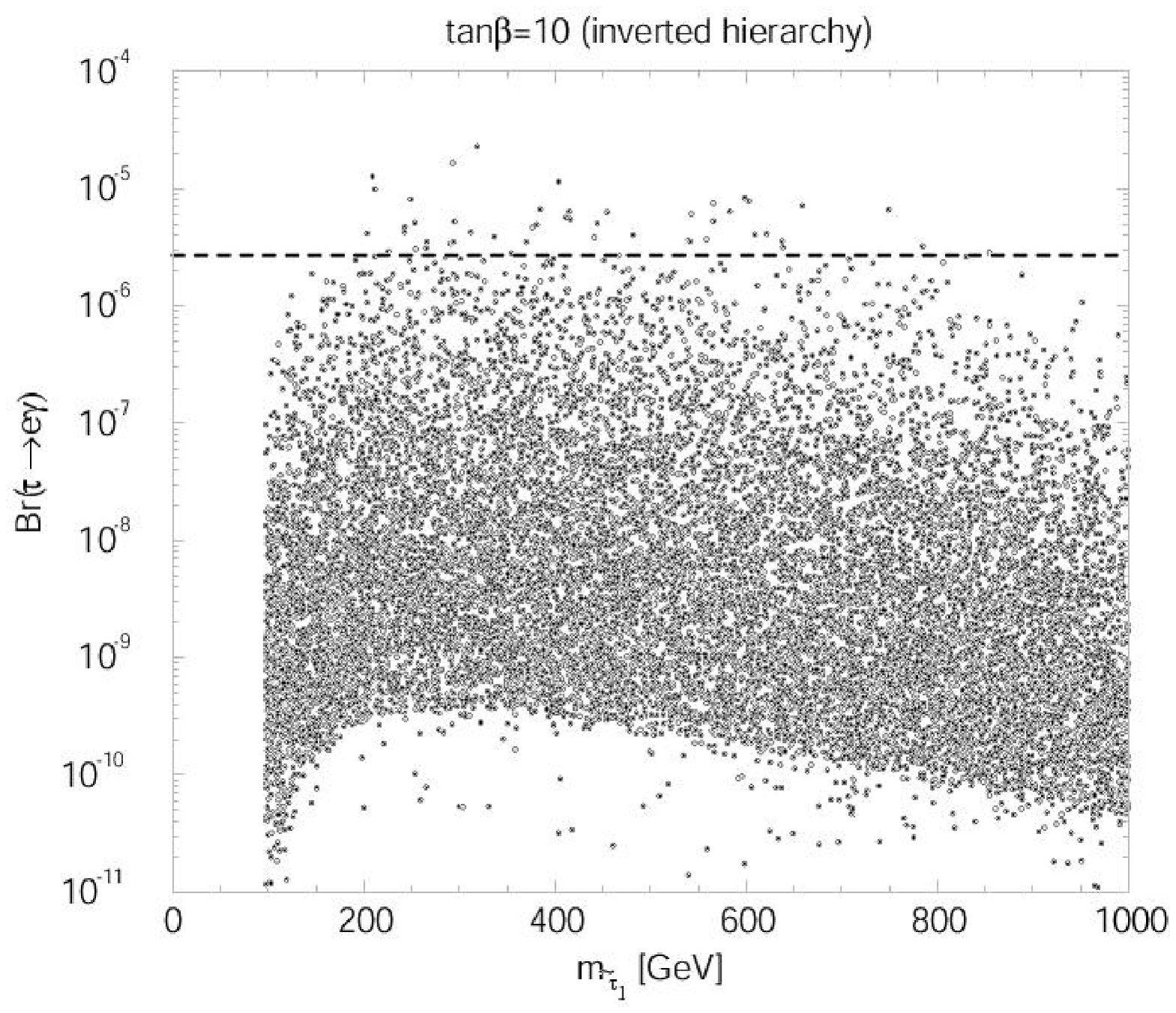} }
\caption{\it 
Scatter plot of Br$(\tau\to e \gamma)$ against the left-handed smuon
mass for the ansatz $H_2$. We assume the normal light-neutrino mass
hierarchy and the inverted hierarchy. The input parameters for the
supersymmetry-breaking parameters are the same as in Fig.~1.
\vspace*{0.5cm}}
\end{figure}

Finally, we comment on the other LFV tau decay processes.  The LFV tau
decay modes to three charged leptons are suppressed in the supersymmetric
models if $R$ parity is conserved. In the almost parameter space the
photonic penguin diagram dominates over the other diagrams.\footnote{
If $\tan\beta$ is very large and pseudoscalar Higgs mass is light, 
the Higgs penguin may give a sizable contribution \cite{Babu:2002et}. 
}
Then, we can get the following approximate formulas,
\begin{eqnarray}
\frac{{\rm Br}(\tau\rightarrow 3\mu)}
    {{\rm Br}(\tau\rightarrow \mu \gamma)}
&\simeq&
\frac{{\rm Br}(\tau\rightarrow e 2\mu )}
    {{\rm Br}(\tau\rightarrow e \gamma)}
\simeq
 \frac{1}{440}\,,
\\
\frac{{\rm Br}(\tau\rightarrow \mu 2 e )}
    {{\rm Br}(\tau\rightarrow \mu \gamma)}
&\simeq&
\frac{{\rm Br}(\tau\rightarrow 3e)}
    {{\rm Br}(\tau\rightarrow e \gamma)}
\simeq
\frac1{94}\,.
\end{eqnarray}
The branching ratio for $\mu e^+ e^-(e e^+ e^-)$ is larger than 
that to $\mu\mu^+\mu^-(e\mu^+\mu^-)$, because the phase space is larger. 
 
\vskip 0.5in
\vbox{
\noindent{ {\bf Acknowledgments} } \\
\noindent  
This work is partially supported by the Grant-in-Aid for Scientific
Research from the Ministry of Education, Science, Sports and Culture
of Japan (No. 13135207 and No. 14046225).}

\end{document}